\documentclass[12pt]{article}
\usepackage{graphics}
\begin{document}

\begin{center}{
{\Large \bf Asymptotic entanglement in open quantum systems}\\
\vskip 0.5truecm
{Aurelian Isar}\\
{Department of Theoretical Physics, National Institute of Physics
and Nuclear Engineering\\
Bucharest-Magurele, Romania
\\
{\it isar@theory.nipne.ro}}
}
\end{center}

\begin{abstract}
In the framework of the theory of open systems based on completely
positive quantum dynamical semigroups,
we solve in the asymptotic long-time regime the master equation for two
independent harmonic oscillators interacting with an environment.
We give a description of the continuous-variable asymptotic
entanglement in terms of the covariance matrix of the considered
subsystem for an arbitrary Gaussian input state.
Using Peres--Simon necessary and sufficient condition for separability of
two-mode Gaussian states, we show that for certain classes of
environments the initial state evolves asymptotically to an
entangled equilibrium bipartite state, while for other values of the
coefficients describing the environment, the asymptotic state is
separable.
We calculate also the logarithmic negativity characterizing the
degree of entanglement of the asymptotic state.
\end{abstract}

\section{Introduction}	

When two systems are immersed in an environment, then, besides
and at the same time with the quantum decoherence phenomenon, the
environment can also generate a quantum entanglement of the
two systems \cite{ben2,ben3}. In certain circumstances, the environment
enhances the entanglement and in others it suppresses the entanglement
and the state describing the two systems becomes separable. The structure of the
environment may be such that not only
the two systems become entangled, but also such that the entanglement
is maintained for a definite time or a certain amount of
entanglement survives in the asymptotic long-time regime.

In the present paper we investigate, in the framework of the theory of open systems
based on completely
positive quantum dynamical semigroups, the existence of the continuous variable
asymptotic entanglement for a subsystem composed of two identical harmonic
oscillators interacting with an
environment.  We are interested in discussing the correlation effect of the
environment, therefore we assume that the two systems are
independent, i.e. they do not interact directly. The initial
state of the subsystem is taken of Gaussian form and the
evolution under the quantum dynamical semigroup assures the
preservation in time of the Gaussian form of the state.
We only investigate here the asymptotic behaviour of the subsystem
states. The time evolution of the entanglement, in particular
the possibility of the so-called "entanglement sudden death",
that is suppression of the entanglement at a certain finite
moment of time, will be discussed in a future work.

The organizing of the paper is as follows.
In Sec. 2 we write the equations of motion in the Heisenberg
picture for two independent harmonic oscillators interacting with a general
environment. With these equations we derive in Sec. 3 the asymptotic values
of variances and
covariances of coordinates and momenta which enter the
asymptotic covariance matrix. Then, by using the
Peres-Simon necessary and sufficient condition for separability
of two-mode Gaussian states \cite{per,sim}, we investigate the
behaviour of the environment induced entanglement in the limit
of long times. We show that for certain classes of environments
the initial state evolves asymptotically to an equilibrium
state which is entangled, while for other values of the
parameters describing the environment, the entanglement is
suppressed and the asymptotic state is separable. The existence
of the quantum correlations between the two systems in the
asymptotic long-time regime is the result of the competition
between entanglement and quantum decoherence. A summary is given in
Sec. 4.

\section{Equations of motion for two independent
harmonic oscillators}

We study the dynamics
of the subsystem composed of two identical non-interacting
oscillators in weak interaction with an environment. In the axiomatic formalism
based on completely positive quantum dynamical semigroups, the irreversible time
evolution of an open
system is described by the following general quantum Markovian master equation
for an operator $A$ (Heisenberg representation) \cite{lin,rev}:
\begin{eqnarray}{dA(t)\over dt}={i\over \hbar}[H,A(t)]+{1\over
2\hbar}\sum_j(V_j^{\dagger}[A(t),
V_j]+[V_j^{\dagger},A(t)]V_j).\label{masteq}\end{eqnarray}
Here, $H$ denotes the Hamiltonian of the open system
and $V_j, V_j^\dagger$ which are operators defined on the Hilbert space of $H,$
represent the interaction of the open system
with the environment. Being interested
in the set of Gaussian states, we introduce those quantum
dynamical semigroups that preserve that set. Therefore $H$ is
taken to be a polynomial of second degree in the coordinates
$x,y$ and momenta $p_x,p_y$ of the two quantum oscillators and
$V_j,V_j^{\dagger}$ are taken polynomials of only first degree
in these canonical observables. Then in the linear space
spanned by the coordinates and momenta there exist only four
linearly independent operators $V_{j=1,2,3,4}$ \cite{san}
: \begin{eqnarray}
V_j=a_{xj}p_x+a_{yj}p_y+b_{xj}x+b_{yj}y,\end{eqnarray} where
$a_{xj},a_{yj},b_{xj},b_{yj}\in {\bf C}.$
The Hamiltonian $H$ of the two uncoupled identical harmonic
oscillators of mass $m$ and frequency $\omega$ is given by \begin{eqnarray}
H={1\over 2m}(p_x^2+p_y^2)+{m\omega^2\over
2}(x^2+y^2).\end{eqnarray}

The fact that the evolution is given by a dynamical semigroup
implies the positivity of the following matrix formed by the
scalar products of the four vectors $ {\bf a}_x, {\bf b}_x,
{\bf a}_y, {\bf b}_y$ whose entries are the components $a_{xj},b_{xj},a_{yj},b_{yj},$
respectively:
\begin{eqnarray}\frac{1}{2} \hbar \left(\matrix
{{\bf a}_x {\bf a}_x)&({\bf a}_x {\bf b}_x) &({\bf a}_x
{\bf a}_y)&({\bf a}_x {\bf b}_y) \cr ({\bf b}_x {\bf a}_x)&({\bf
b}_x {\bf b}_x) &({\bf b}_x {\bf a}_y)&({\bf b}_x {\bf b}_y)
\cr ({\bf a}_y {\bf a}_x)&({\bf a}_y {\bf b}_x) &({\bf a}_y
{\bf a}_y)&({\bf a}_y {\bf b}_y) \cr ({\bf b}_y {\bf
a}_x)&({\bf b}_y {\bf b}_x) &({\bf b}_y {\bf a}_y)&({\bf b}_y
{\bf b}_y}\right).
\end{eqnarray}
For simplicity we take this matrix of the following form, where
all coefficients $D_{xx}, D_{xp_x},$... and $\lambda$ are real quantities:
\begin{eqnarray} \left(\matrix{D_{xx}&- D_{xp_x} - i \hbar \lambda/2&D_{xy}& -
D_{xp_y} \cr - D_{xp_x} + i \hbar\lambda/2&D_{p_x p_x}&-
D_{yp_x}&D_{p_x p_y} \cr D_{xy}&- D_{y p_x}&D_{yy}&- D_{y p_y}
- i \hbar \lambda/2 \cr - D_{xp_y} &D_{p_x p_y}&- D_{yp_y} + i
\hbar \lambda/2&D_{p_y p_y}}\right).\label{coef} \end{eqnarray} It follows that
the principal minors of this matrix are positive or zero. From
the Cauchy-Schwarz inequality the following relations for the
coefficients defined in Eq. (\ref{coef}) hold (from now on we
put, for simplicity, $\hbar=1$): \begin{eqnarray} D_{xx}D_{yy}-D^2_{xy}\ge
0,~ D_{xx}D_{p_xp_x}-D^2_{xp_x}\ge\frac{\lambda^2}{4},~
D_{xx}D_{p_yp_y}-D^2_{xp_y}\ge 0,~\nonumber \\
D_{yy}D_{p_xp_x}-D^2_{yp_x}\ge 0,~
D_{yy}D_{p_yp_y}-D^2_{yp_y}\ge\frac{\lambda^2}{4},~
D_{p_xp_x}D_{p_yp_y}-D^2_{p_xp_y}\ge 0.\end{eqnarray}

The covariance of operators $A_1$ and
$A_2$ can be written with the density operator $\rho$,
describing the initial state of the quantum system, as follows:
\begin{eqnarray}\sigma_{A_1A_2}(t)={1\over 2}{\rm
Tr}(\rho(A_1A_2+A_2A_1)(t)).\end{eqnarray}

We introduce the following $4\times 4$ bimodal covariance matrix:
\begin{eqnarray}\sigma(t)=\left(\matrix{\sigma_{xx}&\sigma_{xp_x} &\sigma_{xy}&
\sigma_{xp_y}\cr \sigma_{xp_x}&\sigma_{p_xp_x}&\sigma_{yp_x}
&\sigma_{p_xp_y}\cr \sigma_{xy}&\sigma_{yp_x}&\sigma_{yy}
&\sigma_{yp_y}\cr \sigma_{xp_y}&\sigma_{p_xp_y}&\sigma_{yp_y}
&\sigma_{p_yp_y}}\right).\label{covar} \end{eqnarray}
By direct calculation we obtain \cite{san} (${\rm T}$ denotes the transposed matrix):
\begin{eqnarray}{d \sigma\over
dt} = Y \sigma + \sigma Y^{\rm T}+2 D,\label{vareq}\end{eqnarray} where
\begin{eqnarray} Y=\left(\matrix{ -\lambda&1/m&0 &0\cr -m\omega^2&-\lambda&0&
0\cr 0&0&-\lambda&1/m \cr 0&0&-m\omega^2&-\lambda}\right),
 ~~D=\left(\matrix{
D_{xx}& D_{xp_x} &D_{xy}& D_{xp_y} \cr D_{xp_x}&D_{p_x p_x}&
D_{yp_x}&D_{p_x p_y} \cr D_{xy}& D_{y p_x}&D_{yy}& D_{y p_y}
\cr D_{xp_y} &D_{p_x p_y}& D_{yp_y} &D_{p_y p_y}} \right).\end{eqnarray}
The time-dependent
solution of Eq. (\ref{vareq}) is given by \cite{san}
\begin{eqnarray}\sigma(t)= M(t)(\sigma(0)-\sigma(\infty)) M^{\rm
T}(t)+\sigma(\infty),\end{eqnarray} where $M(t)=\exp(Yt).$ The matrix
$M(t)$ has to fulfil the condition $\lim_{t\to\infty} M(t)=0.$
In order that this limit exists, $Y$ must only have eigenvalues
with negative real parts. The values at infinity are obtained
from the equation \cite{san} \begin{eqnarray}
Y\sigma(\infty)+\sigma(\infty) Y^{\rm T}=-2 D.\label{covarinf}\end{eqnarray}

\section{Environment induced entanglement}

The two-mode Gaussian state is entirely specified by its
covariance matrix $\sigma$ (\ref{covar}), which is a real,
symmetric and positive matrix with the following block
structure:
\begin{eqnarray}
\sigma=\left(\begin{array}{cc}A&C\\
C^{\rm T}&B \end{array}\right),
\end{eqnarray}
where $A$, $B$ and $C$ are $2\times 2$ matrices. Their entries
are correlations of the canonical operators $x,y,p_x,p_y$, $A$
and $B$ denote the symmetric covariance matrices for the
individual reduced one-mode states, while the matrix $C$
contains the cross-correlations between modes. In the following we shall consider
environments for which $D_{xx}=D_{yy}, D_{xp_x}=D_{yp_y},
D_{p_xp_x}=D_{p_yp_y}, D_{xp_y}=D_{yp_x}.$ Then both unimodal covariance
matrices are equal, $A=B$ and the entanglement matrix $C$ is
symmetric. With the chosen coefficients, we obtain from Eq. (\ref{covarinf}) the
following elements of the asymptotic entanglement matrix $C$:
\begin{eqnarray}\sigma_{xy} (\infty) =
\frac{m^2(2\lambda^2+\omega^2)D_{xy}+2m\lambda
D_{xp_y}+D_{p_xp_y}} {2m^2\lambda(\lambda^2+\omega^2)},\end{eqnarray}
\begin{eqnarray}\sigma_{xp_y}(\infty)=
\sigma_{yp_x}(\infty)=\frac{-m^2\omega^2 D_{xy}+2m\lambda
D_{xp_y}+ D_{p_xp_y}}{2m(\lambda^2+\omega^2)},\end{eqnarray}
\begin{eqnarray}\sigma_{p_xp_y} (\infty) =
\frac{m^2\omega^4D_{xy}-2m\omega^2\lambda D_{xp_y}+(2\lambda^2
+\omega^2)D_{p_xp_y}}{2\lambda(\lambda^2+\omega^2)}.\end{eqnarray} The
elements of matrices $A$ and $B$ are obtained by taking $x=y$ in the previous
expressions. We calculate the determinant of the entanglement matrix and obtain:
\begin{eqnarray} \det
C=\frac{1}{4\lambda^2(\lambda^2+\omega^2)}[(m\omega^2D_{xy}+
\frac{1}{m}
D_{p_xp_y})^2+4\lambda^2(D_{xy}D_{p_xp_y}-D_{xp_y}^2)].\end{eqnarray}
Gaussian states with $\det C\ge 0$ are
separable states, but for $\det C <0$ it may be possible that
the asymptotic equilibrium states are entangled.
In order to investigate whether the environment can actually
entangle the two independent systems, we can use the partial
transposition criterion \cite{per,sim}: a state is
entangled if and only if the operation of partial transposition
does not preserve its positivity. Simon \cite{sim} obtained the
following necessary and sufficient criterion for separability:
$S\ge 0,$ where \begin{eqnarray} S\equiv\det A \det B+(\frac{1}{4} -|\det
C|)^2- {\rm Tr}[AJCJBJC^{\rm T}J]- \frac{1}{4}(\det A+\det B)
\label{sim1}\end{eqnarray} and $J$ is the $2\times 2$ symplectic matrix.

In order to analyze the possible persistence of the environment
induced entanglement in the asymptotic long-time regime, we
consider the environment characterized by the following values
of its parameters: $m^2\omega^2D_{xx}=D_{p_xp_x},~~D_{xp_x}=0,
~m^2\omega^2D_{xy}=D_{p_xp_y}.$ In this case the Simon
expression (\ref{sim1}) takes the form: \begin{eqnarray} S=
\left(\frac{m^2\omega^2(D_{xx}^2-D_{xy}^2)}{\lambda^2}+
\frac{D_{xp_y}^2}{\lambda^2+\omega^2}-\frac{1}{4}\right)^2-4\frac
{m^2\omega^2D_{xx}^2D_{xp_y}^2}{\lambda^2(\lambda^2+
\omega^2)}.\label{sim2}\end{eqnarray} For environments characterized by
such coefficients that the expression (\ref{sim2}) is strictly negative,
the asymptotic final state is entangled. In particular, if
$D_{xy}=0,$ we obtain that $S<0,$ i.e. the asymptotic final
state is entangled, for the following range of values of the
coefficient $D_{xp_y}$ characterizing the environment \cite{arus}:
\begin{eqnarray}
\frac{m\omega
D_{xx}}{\lambda}-\frac{1}{2}<\frac{D_{xp_y}}{\sqrt{\lambda^2
+\omega^2}}<\frac{m\omega
D_{xx}}{\lambda}+\frac{1}{2},\label{insep}\end{eqnarray} where the
coefficient $D_{xx}$ satisfies the condition $m\omega
D_{xx}/\lambda\ge 1/2,$ equivalent with the unimodal
uncertainty relation. If the coefficients do not fulfil the
inequalities (\ref{insep}), then $S\ge 0$ and the
asymptotic state of the considered system is
separable.

\begin{figure}
\includegraphics{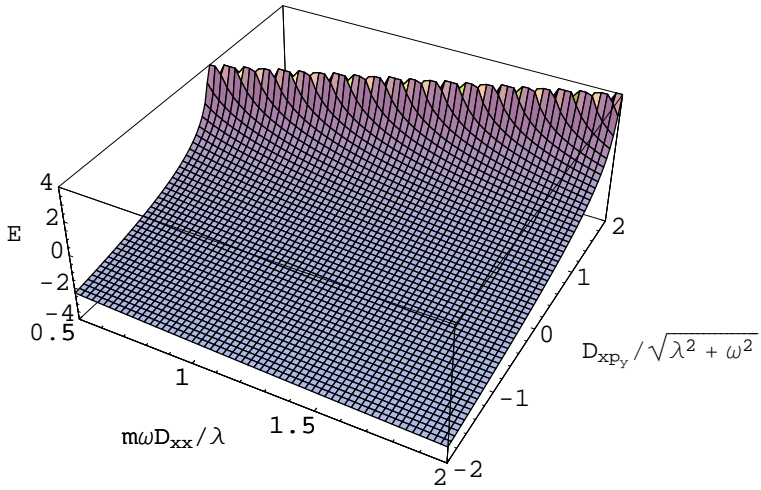}
\caption{Logarithmic negativity versus scaled diffusion coefficients.}
\label{fig:1}       % Give a unique label
\end{figure}

We apply the measure of entanglement based on negative eigenvalues
of the partial transpose of the subsystem density matrix. For a Gaussian density
operator, the negativity is completely defined
by the symplectic spectrum of the partial transpose of the
covariance matrix. The logarithmic negativity $E=-\frac{1}{2}\log_2[4f(\sigma)]$
determines the strength of entanglement for $E>0.$ If $E\le 0,$ then the state is
separable. Here \begin{eqnarray} f(\sigma)=\frac{1}{2}(\det A +\det
B)-\det C-\sqrt{\left[\frac{1}{2}(\det A+\det B)-\det
C\right]^2-\det\sigma}.\end{eqnarray}  In our case the logarithmic negativity
has the
form
\begin{eqnarray} E=-\log_2\left[2\left|\frac{m\omega
D_{xx}}{\lambda}-\frac{D_{xp_y}}{\sqrt{\lambda^2
+\omega^2}}\right|\right].\end{eqnarray}
It depends only on the diffusion and dissipation coefficients
characterizing the environment and does not depend on the initial
Gaussian state. In Fig. 1 logarithmic negativity $E$ is represented versus scaled
diffusion coefficients $ D_{xx}$ and $D_{xp_y}.$

\section{Summary}

In the framework of the theory of open quantum systems based on completely
positive quantum dynamical
semigroups, we investigated
the existence of the asymptotic quantum entanglement for a subsystem
composed of two uncoupled identical harmonic oscillators
interacting with an environment.
By using the Peres-Simon
necessary and sufficient condition for separability of two-mode
Gaussian states, we have shown that for certain classes of
environments the initial state evolves asymptotically to an
equilibrium state which is entangled, i.e. there exist
non-local quantum correlations for the bipartite states of the
two harmonic oscillator subsystem, while for other values of the
coefficients describing the environment, the asymptotic state
is separable. We determined also the logarithmic negativity characterizing the
degree of entanglement of the asymptotic state.
Due to the increased interest manifested towards
the continuous variables approach to quantum information theory, these results,
in particular the
possibility of maintaining a bipartite entanglement in a
diffusive-dissipative environment for asymptotic long
times, might be useful for applications in the field of quantum information
processing and communication.

\section*{Acknowledgments}

The author acknowledges the financial support received within
the Project CEEX 68/2005.

\end{document}